\documentclass[acmtog]{acmart}

\usepackage{booktabs} 

\usepackage{arydshln}

\usepackage{xcolor}
\usepackage{tikz}
\usepackage{multirow}
\usepackage{enumitem}

\usepackage[ruled]{algorithm2e} 

\SetAlFnt{\small}
\SetAlCapFnt{\small}
\SetAlCapNameFnt{\small}
\SetAlCapHSkip{0pt}

\begin{document}
\title{How Does a Virtual Agent Decide Where to Look? Symbolic Cognitive Reasoning for Embodied Head Rotation}

\author{Juyeong Hwang}
\authornote{These authors contributed equally to this work.}
\affiliation{%
  \institution{IIIXR Lab, Korea University}
  \city{Seoul}
  \country{South Korea}
}
\email{05judy02@korea.ac.kr}

\author{Seong-Eun Hong}
\authornotemark[1]
\affiliation{%
  \institution{IIIXR Lab, Korea University}
  \city{Seoul}
  \country{South Korea}
}
\email{seong\_eun@korea.ac.kr}

\author{JaeYoung Seon}
\affiliation{%
  \institution{IIIXR Lab, Kyung Hee University}
  \city{Seoul}
  \country{South Korea}
}
\email{cogongnam@khu.ac.kr}

\author{HyeongYeop Kang}
\authornote{Corresponding author.}
\affiliation{%
  \institution{IIIXR Lab, Korea University}
  \city{Seoul}
  \country{South Korea}
}
\email{siamiz\_hkang@korea.ac.kr}

\renewcommand{\shortauthors}{Hwang, Hong, Seon, and Kang}

\begin{teaserfigure}
    \centering
    \includegraphics[width=0.85\textwidth]{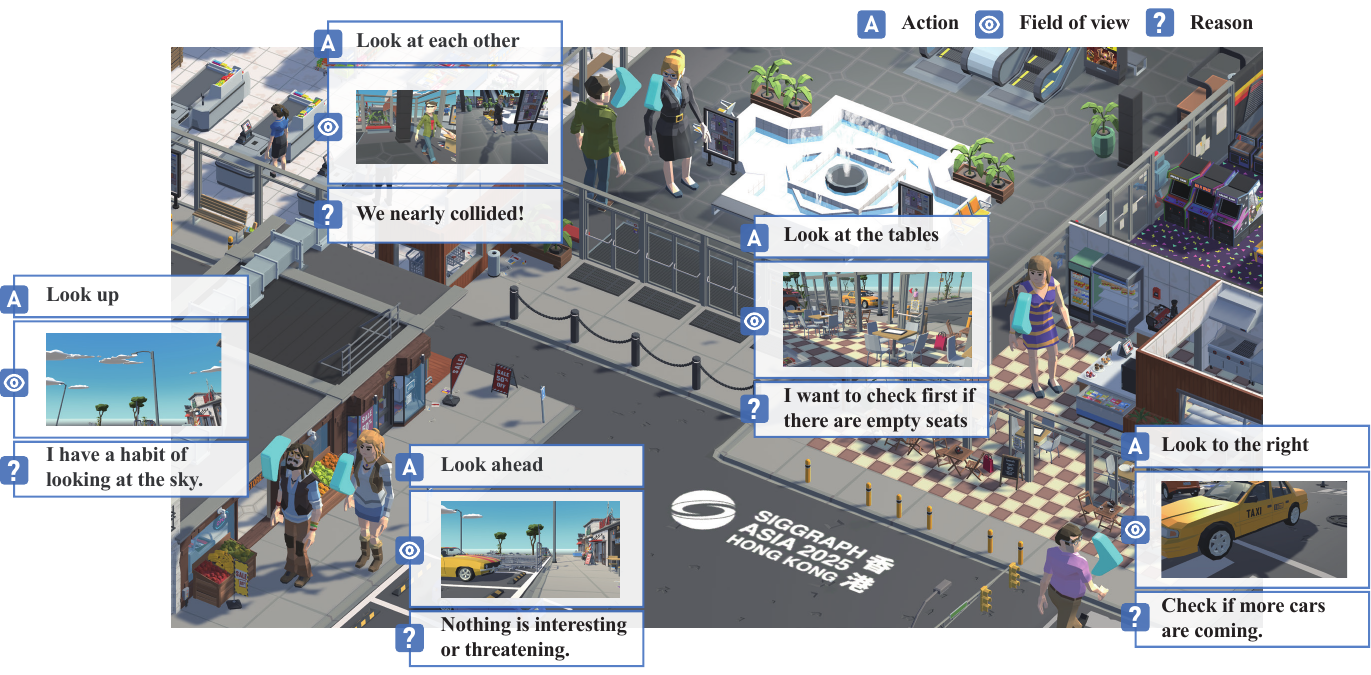}
    \caption{Our system infers not only \textit{where} each avatar should look, but also \textit{why}. In this outdoor–indoor scene, embodied agents decide where to look based on five symbolic motivational drivers---for example, checking oncoming traffic, reading a map, or searching for vacant seats.
    Insets show the first-person view captured at each head orientation, together with the agent’s internal rationale produced by SCORE’s vision–language reasoning pipeline.}
    \label{fig:teaser}
\end{teaserfigure}

\begin{abstract}
Natural head rotation is critical for believable embodied virtual agents, yet this micro-level behavior remains largely underexplored. 
While head-rotation prediction algorithms could, in principle, reproduce this behavior, they typically focus on visually salient stimuli and overlook the cognitive motives that guide head rotation. 
This yields agents that look at conspicuous objects while overlooking obstacles or task-relevant cues, diminishing realism in a virtual environment. 
We introduce \textbf{SCORE}, a \textbf{S}ymbolic \textbf{CO}gnitive \textbf{R}easoning framework for \textbf{E}mbodied Head Rotation, a data-agnostic framework that produces context-aware head movements without task-specific training or hand-tuned heuristics. A controlled VR study ($N$=20) identifies five motivational drivers of human head movements: \textit{Interest}, \textit{Information Seeking}, \textit{Safety}, \textit{Social Schema}, and \textit{Habit}. 
SCORE encodes these drivers as symbolic predicates, perceives the scene with a Vision–Language Model (VLM), and plans head poses with a Large Language Model (LLM). The framework employs a hybrid workflow: the VLM-LLM reasoning is executed offline, after which a lightweight FastVLM performs online validation to suppress hallucinations while maintaining responsiveness to scene dynamics. 
The result is an agent that predicts not only \textit{where} to look but also \textit{why}, generalizing to unseen scenes and multi-agent crowds while retaining behavioral plausibility. 
\end{abstract}

%
%



%
%


\maketitle

\section{Introduction}
Realistic and context-aware virtual agents have long been a central research focus in computer graphics~\cite{lerner2010context, steel2010context,narang2016pedvr, curtis2022toward}. 
One of the key aspects of believable agent behavior lies in natural head rotations that reveal what it notices, what it intends, and how it handles risk. 

Synthesizing such micro-behavior is a long-standing challenge. Yet recent head movement synthesizers still treat attention as a colored blob on the image plane. Trained on saliency maps~\cite{xu2018gaze,zhu2020learning,yang2021hierarchical,rondon2022track}, they steer the camera toward the conspicuous elements while ignoring why a human might look elsewhere. Real pedestrians, for example, repeatedly scan the flow of oncoming cars, bicycles, and walkers, balancing safety, curiosity, and social convention in a single sweep. Saliency-only models over-predict attraction to bright or moving objects and under-predict purposeful reorientations guided by cognitive processes, degrading realism in cluttered, dynamic scenes.

To address these limitations, we propose \textit{SCORE}, a \textbf{S}ymbolic \textbf{CO}gnitive \textbf{R}easoning framework for \textbf{E}mbodied head rotation. 
A VR study with twenty participants across ten scenarios identified five motivational drivers of human head movements: \textit{Interest}, \textit{Information-seeking}, \textit{Safety}, \textit{Social schema}, and \textit{Habit}. \textit{SCORE} exploits these drivers as symbolic predicates and embeds them in a two-stage controller. 

An offline Deliberative Perception–Planning Stage (DPS) combines a Vision–Language Model (VLM) and a Large Language Model (LLM) to draft a time-stamped head-orientation plan. At run time, a lightweight Reactive Execution Stage (RES) validates each pose about two seconds before display with FastVLM; if the live view deviates, RES substitutes a context-appropriate alternative. 
In practice, this ``near-online” correction rejects actions based on hallucinations, absorbs late scene edits, and lets agents respond plausibly to user maneuvers or unforeseen obstacles—all without rerunning the costly VLM–LLM loop. 

In summary, our contributions are as follows: 
\begin{itemize}
 \item \textbf{Cognitive Reasoning for plausible head-motion:} 
 We introduce \emph{SCORE}, the first framework that synthesizes context-aware head orientations via cognitive reasoning. 
 \item \textbf{Zero-shot generalization:} 
 \textit{SCORE} transfers to unseen environments without additional data, tuning, or code changes. 
  \item \textbf{Human-inspired symbolic model:}
Five motivational drivers, extracted from a VR user study, are encoded as predicates that significantly improve behavioural plausibility.
 \item \begin{sloppypar} \textbf{Hybrid control:} 
 A semi-reactive pipeline pairs offline VLM–LLM planning with lightweight FastVLM validation, reducing hallucinations while maintaining responsiveness to scene dynamics.
 \end{sloppypar}
\end{itemize}

\section{RELATED WORK}
\subsection{Head-Movement Prediction}
Head orientation is widely regarded as a reliable proxy for observable visual attention: in social activity or outdoor tasks, where a person turns their head often signals what they are thinking about and concentrating on~\cite{stiefelhagen2002modeling, danninger2005using, ba2008recognizing, nguyen2018your}. 
In other words, predicting those turns requires more than stimulus detection; it calls for an understanding of the underlying cognitive intent.

Head movement prediction has been widely researched to optimize user experience in 360-degree video consumption~\cite{fan2017fixation,aladagli2017predicting}.
Early methods extrapolated head trajectories via linear filters~\cite{qian2016optimizing, duanmu2017prioritized}, but they ignored the link between attention and changing content.
Subsequent work fused saliency with motion history; PanoSalNet + LSTM captured the equatorial bias of HMD viewers and improved accuracy during rapid turns~\cite{nguyen2018your}. TRACK advanced this line by balancing trajectory and saliency cues with a Structural-RNN~\cite{jain2016structural}, achieving state-of-the-art robustness across exploratory and focus-driven clips~\cite{rondon2022track}.


\begin{sloppypar}
Despite these gains, saliency- and trajectory-based predictors remain stimulus-bound; they overlook the cognitive drivers—information seeking, habitual scanning—that govern human re-orientation. 
This gap highlights the importance of believability: beyond mechanical accuracy, motion must express underlying intentions—what classic animation has described as the “illusion of life”~\cite{thomas1995illusion}. In a similar spirit,
\cite{curtis2022toward} argue that plausible motion is tied to higher-level reasoning. Our work follows this direction, using symbolic cognitive predicates and vision--language reasoning to predict not just \emph{where} an agent looks, but \emph{why}.
\end{sloppypar}

\subsection{Vison-Language Reasoning}
Recent advancements in VLM and LLM have enabled their application across various domains, including scene understanding~\cite{fu2024scene}, mathematical problem solving~\cite{liang2023unimath}, and embodied agent control~\cite{kim2024openvla}. 
They increasingly rely on Chain-of-Thought prompting~\cite{wei2022chain}, which forces models to spell out intermediate reasoning~~\cite{yao2023react, shinn2023reflexion}.
By inserting an explicit reasoning layer between perception and action, these systems plan more reliably and generalise across tasks~\cite{song2023llm, yang2023mm}.

We extend this paradigm to virtual-agent head motion, where every turn reflects the agent’s understanding of its goals and surroundings. Unlike saliency or data-driven methods that predict only \emph{where} agents will look, our framework uses VLM–LLM reasoning to infer \emph{why}. 

\section{Motivation and Problem Statement}
Research on realistic virtual-agent behavior has traditionally focused on \emph{macro-level} phenomena, such as crowd simulations~\cite{panayiotou2022ccp, charalambous2023greil, ji2024text} or trajectory prediction~\cite{yue2022human, guo2022end, lin2025progressive, wong2022view, mangalam2021goals}. 
By comparison, \emph{micro-level} actions such as head-orientation remain under-explored, even though subtle head shifts strongly influence perceived plausibility in interactive VR and game settings.

Synthesizing natural head turns is challenging because the motion arises from the integration of scene perception, cognitive evaluation, and decision-making. 
Rather than explicitly modeling these intricate processes, recent studies typically adopt data-driven approaches or visual-saliency proxies~\cite{rondon2022track, nguyen2018your}.
Although such approaches capture conspicuous cues, their shallow treatment of cognition limits generalizability: agents fail to adapt when salient features conflict with task context or social norms, leading to perceptually implausible behaviour in environments.

To overcome these limitations, we first conduct a user study to reveal the underlying rationale behind human head-rotation decisions. Drawing on these insights, we introduce a VLM/LLM–based framework to simulate this decision-making process, enabling robust, data-agnostic head-motion synthesis across diverse and previously unseen scenarios.

\section{User Study}
Realistic head motion emerges from an agent’s on-the-fly interpretation of visual context and task demands.  
To replicate such behaviour, we first require an empirical account of \emph{how} and \emph{why} people rotate their heads while navigating immersive environments.  
The user study, therefore, records human head trajectories in VR together with participants’ own explanations, furnishing both kinematic patterns and the cognitive rationales that produced them.  
These data subsequently guide the design and validation of our VLM/LLM–driven decision model.

\begin{figure}[t]
  \centering
  \includegraphics[width=1\linewidth]{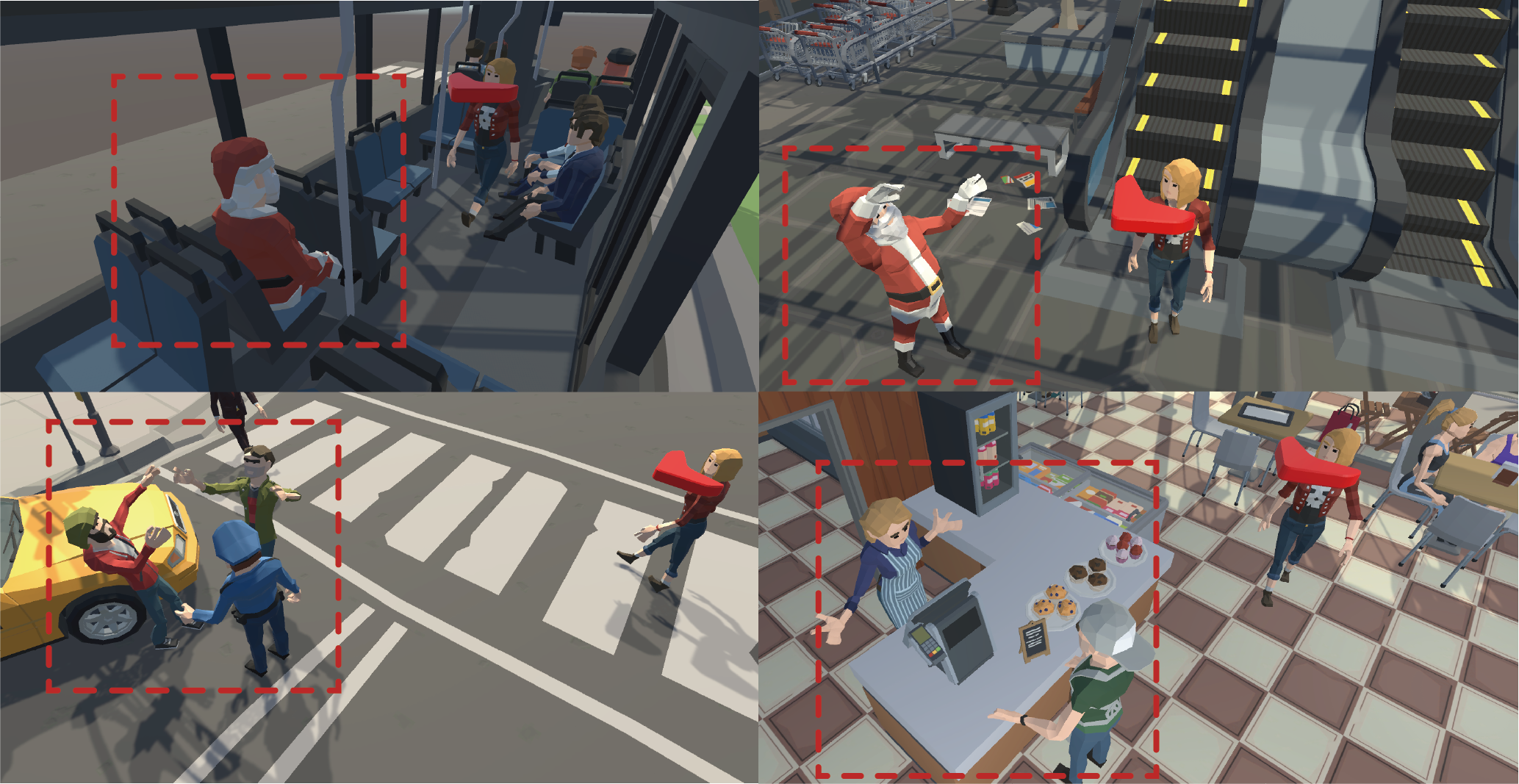}
  \caption{Red dashed boxes highlight elements present only in the APC version of each environment: a Santa, a traffic accident, and arguing people. The corresponding MDC scenes omit such distracting events.}
  \label{fig:APC SceneExample}
\end{figure}

\subsection{Apparatus and Settings}
The study involved 20 participants (11 males, 9 females; $24.27\pm2.60$ yrs), and all participants had prior experience with VR.
We used an Oculus Quest 2 headset driven by a PC (RTX~3090, AMD Ryzen 7 3800XT), paired with controllers.

Five everyday environments (\textit{bus}, \textit{café}, \textit{crosswalk}, \textit{shopping mall}, and \textit{street}) were used for test environments.
To collect diverse data, we introduced two distinct experimental conditions:
\begin{enumerate}[label=\Roman*., topsep=3pt, leftmargin=2em]
\item Minimal-Distraction Condition (MDC): A baseline version of each scene in which all objects are static, free of abrupt motion, or high-contrast coloring. 
\item Attention-Provoking Condition (APC): An augmented version of the same layout that embeds distractors: unexpected obstacles, or motion-triggered distractions. Illustrative examples are provided in ~\autoref{fig:APC SceneExample}.
\end{enumerate}

\begin{figure}[t]
  \centering
  \includegraphics[width=1\linewidth]{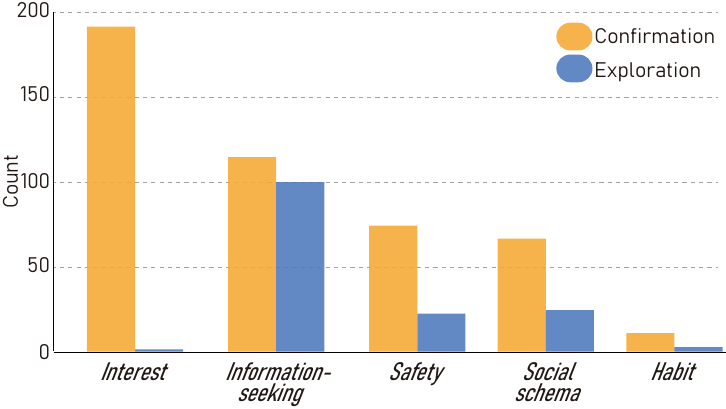}
  \caption{Categorized distribution of participants’ self-reported head-movement rationales.}
  \label{fig:MDC and APC}
\end{figure}

\begin{figure*}[t]
  \centering
  \includegraphics[width=\textwidth]{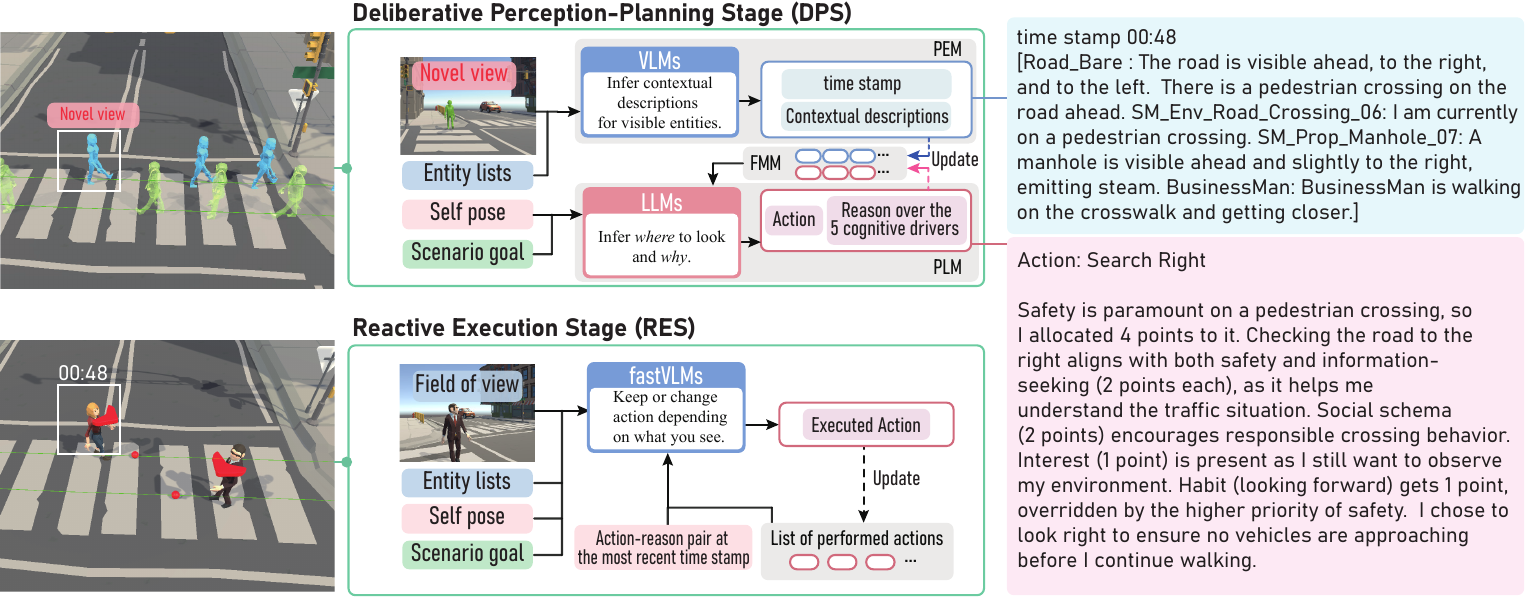}
  \caption{ SCORE pipeline. The two-stage architecture combines offline deliberation with near-online refinement. In the DPS, PEM passes each ``novel” view, the entity lists, and the scenario goal to a VLM, which generates contextual descriptions stored in FMM. The Planning Module then invokes an LLM that reasons over the five cognitive drivers to select an action–reason pair that schedules the next head orientation. During runtime, RES evaluates a 2-second look-ahead image with a lightweight FastVLM; if the pre-planned action is inconsistent with the live context, it is replaced before execution.}
  \label{fig:Architecture}
\end{figure*}

\subsection{Method and Procedure}
A user study was conducted with two experimenters. Upon arrival, each participant provided informed consent and demographic data, then completed a 15-minute orientation (5-minute briefing and up to 10 minutes of free navigation in a neutral city scene). Simulator Sickness Questionnaire~\cite{kennedy1993simulator} (SSQ) scores taken before and after the study showed no significant change, and Simulation Task Load Index~\cite{harris2020development} (SIM-TLX) ratings confirmed low mental workload. 

Each participant performed ten 60-second trials covering all combinations of two attentional settings—MDC and APC—and five everyday environments. In each scenario, participants were instructed to complete a scenario-specific goal: \textit{Bus-}find and sit in an empty seat at the back; \textit{Café-}locate and sit at a table by the window; \textit{Crosswalk-}cross safely to the other side; \textit{Shopping Mall-}move from one end of the mall to the opposite side; and \textit{Street-}walk safely to the far end of the street. A 180-second break was available between trials; trial order was randomized to mitigate sequence effects.

We recorded egocentric video, head-orientation quaternions, and sensor data during each scenario.
Upon completion of all scenarios, participants reviewed their footage and articulated the underlying motive for every head turn. 
To facilitate reliable articulation, participants were presented with example prompts such as ``\textit{it caught my eye,}" or ``\textit{I needed to confirm something.}”

\subsection{Derivation of Motivational Drivers}
Each head-motion trace was first partitioned into \textit{confirmation turns}, minor intrafoveal shifts used to verify peripheral cues, and \textit{exploration turns}, larger reorientations that probe regions outside the current view. 
With these classes, movements were further divided into \textit{object-directed} and \textit{information-seeking}.
These separations clarified whether a head movement was a check of something already noticed or a deliberate search for new information.

We annotated each trajectory and its 200 accompanying participant explanations. This began with a codebook containing the three motivational factors derived from the attention research---\textit{Safety}, \textit{Interest}, and \textit{Information Seeking}~\cite{yang2024interpreting, witchel2016non, bromberg2020neural}. During iterative review, we observed recurrent behaviours that these labels could not capture, such as surveying the general flow of pedestrians in a shopping mall or tilting the head upward to scan the sky when walking. This led us to define two additional categories, \textit{Social Schema} and \textit{Habit}. 
The resulting five-factor set constitutes the motivational taxonomy that supports SCORE’s decision model. \autoref{fig:MDC and APC} visualises how often each motivation category appeared in participant explanations across the various scenarios.

\section{SCORE Framework}
\textit{SCORE} operates through a two-stage pipeline. Deliberative Perception–Planning Stage is the offline planning stage that predicts and schedules the agent’s desired head orientation along the agent's trajectory. Reactive Execution Stage is the near-online validation stage that refines the pre-planned desired head orientation during runtime.    
The overall pipeline is shown in~\autoref{fig:Architecture}; implementation details appear in the supplementary.

\subsection{Deliberative Perception–Planning Stage (DPS)}
DPS runs once for an entire trajectory. At every 0.2s, it submits the agent's current field of view (FOV) to the Perception Module (PEM). 
To avoid redundant processing, PEM compares the incoming view image with the most recent ``novel" view and flags it as novel only if the Structural Similarity Index Measure (SSIM)~\cite{wang2004image} between them is below 60~\cite{yedla2020real}.
For each novel view, a forward ray-tracing pass enumerates visible objects and agents. The novel view image, the resulting entity lists, and the scenario goal are then passed to the VLM, which produces contextual descriptions for every entity. 
These lists and descriptions, stamped with the capture time, are stored in the Foundational Memory Module (FMM).

Afterward, the Planning Module (PLM) invokes an LLM to infer an action–reason pair: the action specifies the target head orientation, and the reason encodes the cognitive motivation.
The LLM receives three sources of information: 1) the scenario goal, 2) the agent’s self-pose, and 3) the current FMM contents. Drawing on these, the LLM performs reasoning over the five empirically derived cognitive drivers- \textit{Interest}, \textit{Information Seeking}, \textit{Safety}, \textit{Social Schema}, and \textit{Habit}- to infer an action-reason pair. The pair is appended to the FMM, enabling subsequent decisions to respect behavioral consistency.

DPS simulates the head turn toward the target at $36^{\circ}/sec$~\cite{grossman1988frequency}.
During the turn, PEM is paused. Once the target pose is achieved, the agent holds its head for 1-2 seconds $\sim N(1.5,0.25^2)$.
The head then returns to the walking direction, mimicking the human tendency to realign the head after a temporal goal (e.g., threat assessment or curiosity) has been met. PEM resumes after the first half of the head's hold period, ready to identify the next novel view. 

Each FMM entry contains the time stamp, the current object and agent lists, and the associated action–reason pair. To maintain robust performance over long horizons, the FMM retains at most twenty entries: ten are the most recent, and the remainder are those deemed most relevant to the scenario goal according to an LLM-based relevance \textit{SCORE}, following recent agent-memory strategies~\cite{seo2025reveca}.  

All hyperparameters—capture frequency, novelty threshold, and memory size—were tuned empirically to produce plausible motions across the ten test scenarios introduced in the user study.

\subsection{Reactive Execution Stage (RES)}
RES operates continuously during the simulation. Every 0.2s, it assembles the agent's current FOV, self-pose, entity list, the scenario goal, the most recent action-reason pair, and the list of actions that have actually been performed, and submits this bundle to a lightweight FastVLM. 
FastVLM determines whether the pre-planned pose remains valid or should be revised.
Executed action is logged via the manage schema used in FMM for ensuring dynamic adaptability.

Because FastVLM still requires inference time, RES queries it against a view predicted two seconds ahead of the display frame. This look-ahead window, much shorter than the full VLM–LLM pass yet sufficient for FastVLM, lets \textit{SCORE} operate in a semi-reactive mode that balances latency with scene dynamics.
Note that the average inference latencies are $3.50\pm0.64$ (VLM), $7.10\pm1.22$ (LLM), and $1.49\pm0.41$ (FastVLM).

\section{Evaluation}
In our evaluation, we use Gemini 1.5 Pro as the VLM and LLM for DPS, and Gemini 1.5 Flash~\cite{team2024gemini} as the FastVLM for RES. To assess \textit{SCORE}’s ability to replicate human-like head-motion patterns, we benchmark it against the saliency-driven baseline \textit{Track}~\cite{rondon2022track} and the user-generated motion data collected in the user study (hereafter, \textit{Human}). 
Similar to prior work that classifies nods and shakes from orientation traces~\cite{switonski2019dynamic, li2024head}, we employ Dynamic Time Warping (DTW)~\cite{sakoe1978dynamic} to compare each simulated quaternion sequence with its human counterpart. 
DTW is well-suited to this task: it aligns two time series sample-by-sample while accommodating local speed variations, yielding an interpretable measure of how closely a generated trajectory follows the spatial path taken by real users~\cite{cuturi2017soft, lerogeron2023approximating}. 
However, because human trajectories vary considerably across individuals even under identical scenarios, due to personal priorities and habits, no single trajectory can serve as a definitive ground truth. To address this, we compare each generated trajectory against all human trajectories collected for the same scenario and report the averaged DTW distance as a similarity score.

At the same time, DTW evaluates trajectory-level alignment but does not capture the perceived believability of the behavior. To complement this limitation, we conducted a small user study to assess realism and plausibility. The outcomes of this study are provided in the supplementary material.

\begin{table}[t]
  \centering
  \caption{DTW comparison between \textit{IntraHuman}, \textit{Track}, and \textit{SCORE} in single-agent settings across five scenarios under MDC and APC conditions. Lower DTW scores indicate closer alignment with human motion data. The best-performing scores are highlighted in bold. The reported $\pm$ values denote 95\% confidence intervals.}
  \label{tab:DTW_QuantitativeComparison}
  \begin{tabular*}{0.475\textwidth}{@{\extracolsep{\fill}}llcc}
    \toprule
    \multirow{2}{*}{Scenario} & \multirow{2}{*}{Method} 
    & \multicolumn{2}{c}{DTW $\downarrow$} \\
    \cmidrule(lr){3-4}
    & & MDC & APC \\
    \midrule
    \multirow{3}{*}{\textit{Bus}}
      & \textit{IntraHuman} & \textit{0.2845$^{\pm .0259}$} & \textit{0.3289$^{\pm .0294}$} \\
      & \textit{Track} & 0.4733$^{\pm .0233}$ & 0.4518$^{\pm .0186}$ \\
      & \textit{SCORE} (Ours) & \textbf{0.4591}$^{\pm .0243}$ & \textbf{0.4029}$^{\pm .0249}$ \\
    \midrule
    \multirow{3}{*}{\textit{Café}}
      & \textit{IntraHuman} & \textit{0.3195$^{\pm .0246}$} & \textit{0.3338$^{\pm .0261}$} \\
      & \textit{Track} & 0.6367$^{\pm .0200}$ & 0.6550$^{\pm .0307}$ \\
      & \textit{SCORE} (Ours) & \textbf{0.4234}$^{\pm .0251}$ & \textbf{0.5072}$^{\pm .0283}$ \\
    \midrule
    \multirow{3}{*}{\textit{Crossing}}
      & \textit{IntraHuman} & \textit{0.1709$^{\pm .0174}$} & \textit{0.1919$^{\pm .0225}$} \\
      & \textit{Track} & 0.5611$^{\pm .0213}$ & 0.6504$^{\pm .0223}$ \\
      & \textit{SCORE} (Ours) & \textbf{0.2760}$^{\pm .0191}$ & \textbf{0.4662}$^{\pm .0241}$ \\
    \midrule
    \multirow{3}{*}{\textit{Mall}}
      & \textit{IntraHuman} & \textit{0.5051$^{\pm .0202}$} & \textit{0.2819$^{\pm .0296}$} \\
      & \textit{Track} & 0.8337$^{\pm .0355}$ & 0.8284$^{\pm .0183}$ \\
      & \textit{SCORE} (Ours) & \textbf{0.5405}$^{\pm .0299}$ & \textbf{0.4569}$^{\pm .0303}$ \\
    \midrule
    \multirow{3}{*}{\textit{Street}}
      & \textit{IntraHuman} & \textit{0.2491$^{\pm .0205}$} & \textit{0.2984$^{\pm .0248}$} \\
      & \textit{Track} & 0.5999$^{\pm .0340}$ & 0.5912$^{\pm .0335}$ \\
      & \textit{SCORE} (Ours) & \textbf{0.4705}$^{\pm .0298}$ & \textbf{0.3778}$^{\pm .0397}$ \\
    \bottomrule
  \end{tabular*}
\end{table}

\begin{table}[t]
  \centering
  \caption{DTW comparison between \textit{Track}, and \textit{SCORE} in multi-agent settings. The best-performing scores are highlighted in bold. The reported $\pm$ values denote 95\% confidence intervals.}
  \label{tab:DTW_QuantitativeComparison_MultiAgent}
  \begin{tabular*}{0.475\textwidth}{@{\extracolsep{\fill}}llc}
    \toprule
    Scenario & Method & DTW $\downarrow$ \\
    \midrule
    \multirow{2}{*}{\textit{Zara01}}
      & \textit{Track} & 0.5632$^{\pm .0255}$ \\
      & \textit{SCORE} (Ours) & \textbf{0.3616}$^{\pm .1015}$ \\
    \midrule
    \multirow{2}{*}{\textit{Zara02}}
      & \textit{Track} & 0.4325$^{\pm .0231}$ \\
      & \textit{SCORE} (Ours) & \textbf{0.2657}$^{\pm .0832}$ \\
    \midrule
    \multirow{2}{*}{\textit{students003}}
      & \textit{Track} & 0.5621$^{\pm .0311}$ \\
      & \textit{SCORE} (Ours) & \textbf{0.4471}$^{\pm .0956}$ \\
    \bottomrule
  \end{tabular*}
\end{table}

\subsection{Single-Agent Benchmark}
\subsubsection{Method}
\label{section:DTWMethod}
We evaluated \textit{SCORE} on the five distinct virtual scenarios—\textit{Bus}, \textit{Café}, \textit{Crossing}, \textit{Mall}, and \textit{Street}—under both MDC and APC conditions, as used in the user study. For each scenario–condition pair, we generated five head-motion trials with \textit{SCORE} and with \textit{Track}, using the same agent body trajectories that were recorded from the user study and replaying them at a constant walking speed of $1.3m/s$~\cite{bohannon1997comfortable}.
All hyperparameters and VLM/LLM prompts were held fixed across scenarios, ensuring a fair comparison.

Head orientation was represented by unit quaternions. The instantaneous angular error between two orientations \(q_1\) and \(q_2\) was computed as 
\begin{equation}
    d(q_1, q_2) = 2 \cdot \arccos\left(\left| \text{dot}(q_1, q_2) \right|\right),
\end{equation}
where \( \text{dot}(q_1, q_2) \) denotes the dot product of the quaternions. This instantaneous error served as the local cost function for DTW. DTW then temporally warped one trajectory with respect to the other to identify the alignment that minimizes the accumulated angular error, compensating for local speed differences between the human and simulated motions. The resulting cost was normalised by the mean sequence length to yield a length-invariant similarity score.

Note that \textit{Track} was designed for 360-degree video, predicting salient gaze over a $360^{\circ}$ equirectangular image.
Applied to first-person navigation with a narrow task-driven FOV, this global-saliency bias is disadvantageous. For a fair comparison, we supply \textit{Track} with only the agent-centred FOV image, and our internal tests confirm that this restriction does not impair its accuracy.

\subsubsection{Results}

\autoref{tab:DTW_QuantitativeComparison} provides the normalized DTW scores for both the \textit{Track} and \textit{SCORE}, with lower values denoting closer similarity to the \textit{Human} head-rotation.
To provide an upper bound performance, we compute \textit{IntraHuman} baseline.
The \textit{Human} is randomly partitioned into two non-overlapping subsets of equal size, and DTW is evaluated between them. Repeating this split ten times and averaging the results yields an empirical lower bound that no model can surpass. 

In every scenario, \textit{SCORE} outperforms the saliency baseline \textit{Track}, confirming that first-person context combined with language-based reasoning yields more human-like trajectories.

The advantage is most pronounced in the \textit{Mall} and \textit{Crosswalk}, where rapid scene changes and the presence of multiple objects occur. \textit{SCORE} also excels in the \textit{Bus}-APC: it consistently attends to a Santa Claus passenger who captured participants’ attention, showing that its cognitive layer can infer the social salience of atypical objects.

Across all five environments, \textit{SCORE} maintains low DTW error, demonstrating robust generalization to dynamic layouts, multiple attractors, and culturally nuanced cues. These results emphasize the value of integrating vision–language reasoning into head-motion synthesis for realistic, context-aware virtual agents.


\subsubsection{Categorical Analysis of Entity Selection}

\begin{figure}[t]
  \centering
  \includegraphics[width=\linewidth]{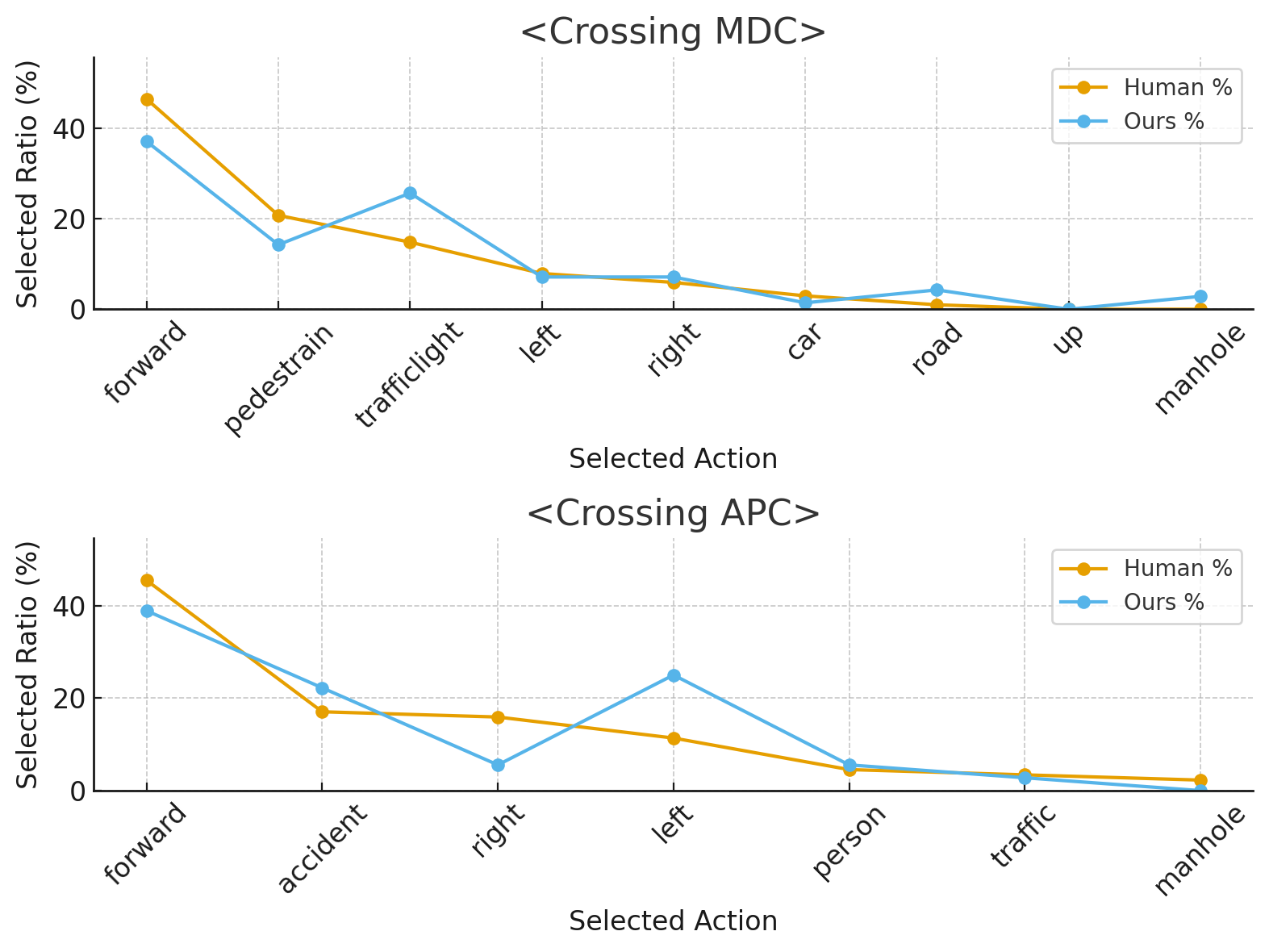}
  \caption{The figure illustrates differences in proportional selections, which indicate prioritization tendencies between humans and the model in the MDC and APC of the \textit{crossing} scene.}

  \label{fig:CategoryAnalysis}
\end{figure}

While DTW evaluation quantifies the similarity of rotational trajectories between humans and the model, it does not reveal whether both selected similar entities within the scene. To complement this, we compared the distribution of entity categories chosen by humans and our \textit{SCORE} model in each scenario (\autoref{fig:CategoryAnalysis}). Under the MDC, humans most frequently selected the forward direction, followed by pedestrians and traffic lights. \textit{SCORE} also prioritized the forward direction but assigned relatively greater weight to traffic lights. We speculate that \textit{SCORE} may exhibit a more ``safety-conscious" behavior by periodically checking traffic lights to ensure safe navigation.

In contrast, under the APC, although the task remained the same, the newly introduced accident scene strongly attracted human attention, leading both humans and the model to allocate substantial attention to this entity. Humans tended to secure safety by scanning both left and right while crossing the street, whereas the agent focused more on the left—the side of the accident—when ensuring safety. Both strategies appear reasonable given the circumstances.

Collectively, the results suggest that both humans and the model turned their gaze for their own valid reasons. The observed differences are less indicative of outright mistakes than of distinct tendencies—or even persona-like variations—in how attention was allocated. See \autoref{fig:CategoryAnalysis} for the full distribution across scenes; detailed results for the \textit{bus},\textit{café}, \textit{mall}, and \textit{street} are provided in the supplementary material.

\subsection{Multi-Agent Benchmark}
\subsubsection{Method}
To evaluate \textit{SCORE}'s generalizability in multi-agent scenarios, we conducted an additional experiment on the UCY crowd dataset~\cite{lerner2007crowds}, which provides frame-level positions, head orientations, and corresponding video footage. 
We reproduced three scenes (\textit{zara01}, \textit{zara02}, and \textit{students003}) in our simulator by replicating the recorded agent configurations and environmental layout. 
The model then generated head-motion trajectories under these conditions, and similarity to the ground-truth orientations was quantified with the DTW metric.

\subsubsection{Results}
\begin{figure*}[t]
  \centering
  \includegraphics[width=1\linewidth]{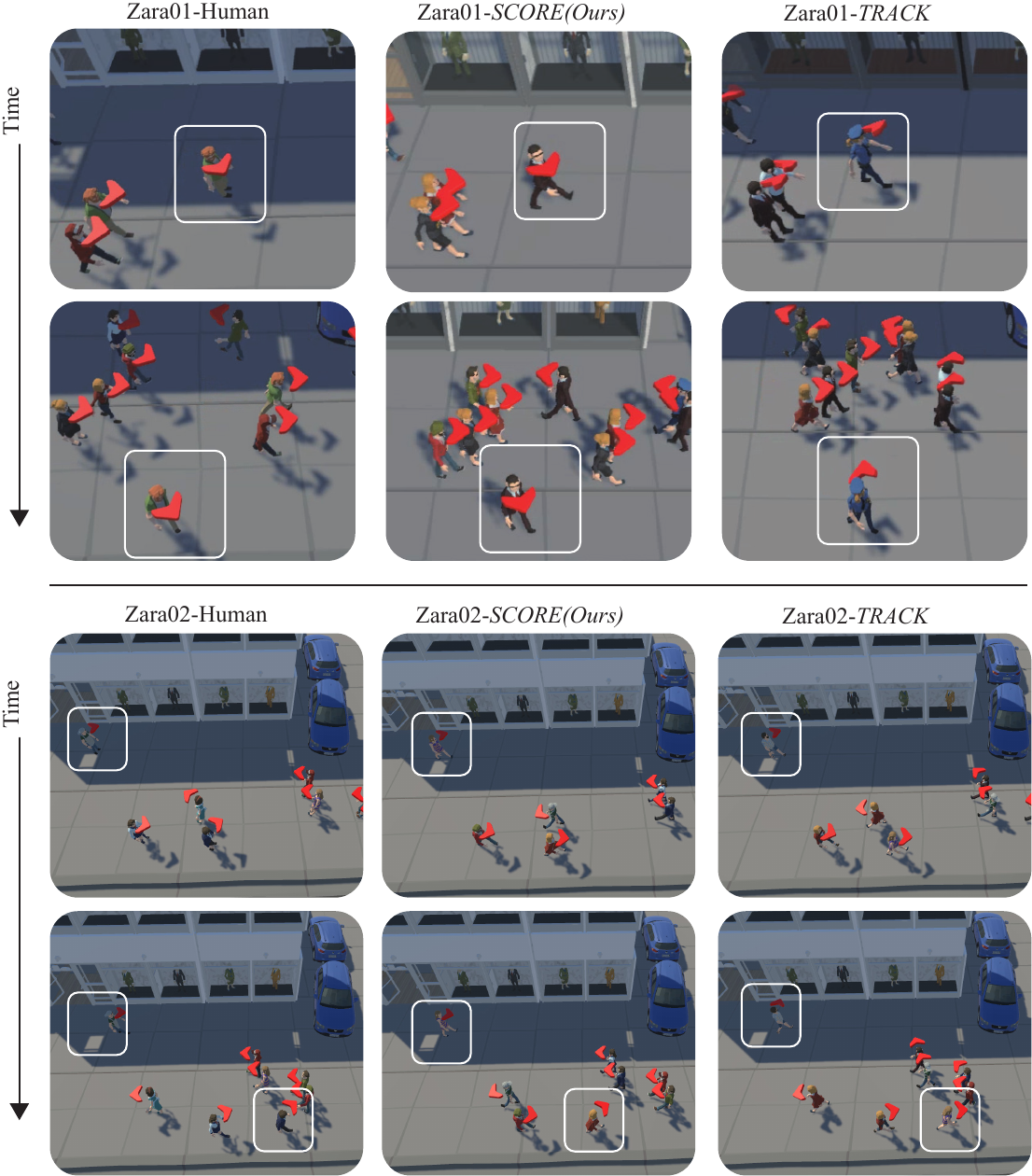}
  \caption{This example illustrates how \textit{SCORE} produces human-like head movements in a multi-agent setting. The highlighted agent attends to the intended walking path, aligning closely with the ground-truth behavior. In contrast, \textit{Track} fixates excessively on static background elements (e.g., storefronts), failing to handle multiple factors (e.g., secure path, understand pedestrian flow) and resulting in implausible head rotations.}
  \label{fig:Multi-agentExample}
\end{figure*}

\textit{SCORE} consistently outperforms the \textit{Track} across all scenes, confirming that its performance scales reliably to multi-agent settings.
Qualitatively, the generated head rotations also align closely with human intuition
In \textit{zara01}, the SCORE agents spontaneously attend to a shop mannequin, mirroring shoppers’ tendency to glance at salient storefront displays. 
In \textit{zara02}, the agents shift their head to the new walking direction before turning, much as real pedestrians scan the path ahead when changing course. 
In \textit{students003}, the agents detect newcomers entering from multiple directions and distribute attention accordingly, reproducing the adaptive scanning seen in crowded situations.
Visual examples are provided \autoref{fig:Multi-agentExample}.

\begin{table}[t]
  \centering
  \caption{Responsiveness benchmark with unseen distractors. Any improvement of \textit{MDC-APC} over \textit{MDC-APC} reflects RES’s ability to detect and adapt to novel stimuli not available at planning time.}
  \label{tab:DTW_Responsiveness Benchmark}
  \begin{tabular*}{0.475\textwidth}{@{\extracolsep{\fill}}lccc}
    \toprule
    \multirow{2}{*}{Scenario} & \multicolumn{3}{c}{DTW $\downarrow$} \\
    \cmidrule(lr){2-4}
    & \textit{MDC-MDC} & \textit{MDC-APC} & \textit{APC-APC} \\
    \midrule
    \textit{Bus}      & 0.4330$^{\pm .0916}$ & 0.4182$^{\pm .0652}$ & 0.4029$^{\pm .0249}$ \\
    \textit{Café}     & 0.5505$^{\pm .1495}$ & 0.5255$^{\pm .0877}$ & 0.5072$^{\pm .0283}$ \\
    \textit{Crossing} & 0.4961$^{\pm .0723}$ & 0.4832$^{\pm .0555}$ & 0.4662$^{\pm .0241}$ \\
    \textit{Mall}     & 0.4909$^{\pm .0728}$ & 0.4724$^{\pm .0598}$ & 0.4569$^{\pm .0303}$ \\
    \textit{Street}   & 0.4875$^{\pm .0642}$ & 0.4322$^{\pm .0601}$ & 0.3778$^{\pm .0397}$ \\
    \bottomrule
  \end{tabular*}
\end{table}

\begin{table}[ht]
  \centering
  \caption{Ablation study of \textit{SCORE} in single-agent settings across five scenarios under MDC and APC conditions.}
  \resizebox{\columnwidth}{!}{\begin{tabular}{llcc}
    \toprule
    \multirow{2}{*}{Scenario} & \multirow{2}{*}{Method} 
    & \multicolumn{2}{c}{DTW $\downarrow$} \\
    \cmidrule(lr){3-4}
    & & MDC & APC \\
    \midrule
    \multirow{9}{*}{\textit{Bus}}
      & \textit{SCORE} & \textbf{0.4591}$^{\pm .0243}$ & \textbf{0.4029}$^{\pm .0249}$ \\
      & \textit{w/o all Motivational Drivers} & 0.4868$^{\pm .0299}$ & 0.5770$^{\pm .0293}$ \\
      & \textit{w/o Interest driver} & 0.4622$^{\pm .0225}$ & 0.4571$^{\pm .0392}$ \\
      & \textit{w/o Information-seeking driver} & 0.4621$^{\pm .0217}$ & 0.4771$^{\pm .0391}$ \\
      & \textit{w/o Safety driver} & 0.4713$^{\pm .0267}$ & 0.4176$^{\pm .0261}$ \\
      & \textit{w/o Social Schema driver} & 0.4515$^{\pm .0315}$ & 0.4871$^{\pm .0297}$ \\
      & \textit{w/o Habit driver} & 0.4719$^{\pm .0210}$ & 0.4470$^{\pm .0210}$ \\
      & \textit{w/o LLM} & 0.4681$^{\pm .0289}$ & 0.4864$^{\pm .0363}$ \\
      & \textit{w/o RES} & 0.5594$^{\pm .0251}$ & 0.4202$^{\pm .0283}$ \\
    \midrule
    \multirow{9}{*}{\textit{Café}}
      & \textit{SCORE} & \textbf{0.4234}$^{\pm .0251}$ & \textbf{0.5072}$^{\pm .0283}$ \\
      & \textit{w/o all Motivational Drivers} & 0.6024$^{\pm .0330}$ & 0.5625$^{\pm .0318}$ \\
      & \textit{w/o Interest driver} & 0.4521$^{\pm .0255}$ & 0.5141$^{\pm .0391}$ \\
      & \textit{w/o Information-seeking driver} & 0.4736$^{\pm .0267}$ & 0.5570$^{\pm .0333}$ \\
      & \textit{w/o Safety driver} & 0.4701$^{\pm .0259}$ & 0.5166$^{\pm .0290}$ \\
      & \textit{w/o Social Schema driver} & 0.4417$^{\pm .0227}$ & 0.5270$^{\pm .0298}$ \\
      & \textit{w/o Habit driver} & 0.4320$^{\pm .0260}$ & 0.5161$^{\pm .0290}$ \\
      & \textit{w/o LLM} & 0.4726$^{\pm .0309}$ & 0.5303$^{\pm .0417}$ \\
      & \textit{w/o RES} & 0.6517$^{\pm .0251}$ & 0.5798$^{\pm .0283}$ \\
    \midrule
    \multirow{9}{*}{\textit{Crossing}}
      & \textit{SCORE} & \textbf{0.2760}$^{\pm .0191}$ & \textbf{0.4662}$^{\pm .0241}$ \\
      & \textit{w/o all Motivational Drivers} & 0.5976$^{\pm .0839}$ & 0.5559$^{\pm .0249}$ \\
      & \textit{w/o Interest driver} & 0.3321$^{\pm .0226}$ & 0.4772$^{\pm .0291}$ \\
      & \textit{w/o Information-seeking driver} & 0.3021$^{\pm .0227}$ & 0.4871$^{\pm .0263}$ \\
      & \textit{w/o Safety driver} & 0.4700$^{\pm .0249}$ & 0.5536$^{\pm .0235}$ \\
      & \textit{w/o Social Schema driver} & 0.3147$^{\pm .0515}$ & 0.4899$^{\pm .0217}$ \\
      & \textit{w/o Habit driver} & 0.2819$^{\pm .0220}$ & 0.4771$^{\pm .0255}$ \\
      & \textit{w/o LLM} & 0.3793$^{\pm .0268}$ & 0.5955$^{\pm .0256}$ \\
      & \textit{w/o RES} & 0.5601$^{\pm .0251}$ & 0.4953$^{\pm .0283}$ \\
    \midrule
    \multirow{9}{*}{\textit{Mall}}
      & \textit{SCORE} & \textbf{0.5405}$^{\pm .0299}$ & \textbf{0.4569}$^{\pm .0303}$ \\
      & \textit{w/o all Motivational Drivers} & 0.5820$^{\pm .0433}$ & 0.6434$^{\pm .0276}$ \\
      & \textit{w/o Interest driver} & 0.5721$^{\pm .0234}$ & 0.6570$^{\pm .0381}$ \\
      & \textit{w/o Information-seeking driver} & 0.5551$^{\pm .0251}$ & 0.4802$^{\pm .0280}$ \\
      & \textit{w/o Safety driver} & 0.5601$^{\pm .0355}$ & 0.4613$^{\pm .0333}$ \\
      & \textit{w/o Social Schema driver} & 0.5721$^{\pm .0321}$ & 0.5002$^{\pm .0265}$ \\
      & \textit{w/o Habit driver} & 0.5723$^{\pm .0251}$ & 0.4901$^{\pm .0223}$ \\
      & \textit{w/o LLM} & 0.5848$^{\pm .0168}$ & 0.5559$^{\pm .0159}$ \\
      & \textit{w/o RES} & 0.5081$^{\pm .0251}$ & 0.4542$^{\pm .0283}$ \\
    \midrule
    \multirow{9}{*}{\textit{Street}}
      & \textit{SCORE} & \textbf{0.4705}$^{\pm .0298}$ & \textbf{0.3778}$^{\pm .0397}$ \\
      & \textit{w/o all Motivational Drivers} & 0.4754$^{\pm .0251}$ & 0.5409$^{\pm .0300}$ \\
      & \textit{w/o Interest driver} & 0.4741$^{\pm .0224}$ & 0.4999$^{\pm .0495}$ \\
      & \textit{w/o Information-seeking driver} & 0.4722$^{\pm .0315}$ & 0.4752$^{\pm .0351}$ \\
      & \textit{w/o Safety driver} & 0.4713$^{\pm .0260}$ & 0.4352$^{\pm .0200}$ \\
      & \textit{w/o Social Schema driver} & 0.4777$^{\pm .0210}$ & 0.4912$^{\pm .0390}$ \\
      & \textit{w/o Habit driver} & 0.4773$^{\pm .0221}$ & 0.4510$^{\pm .0422}$ \\
      & \textit{w/o} LLM & 0.4775$^{\pm .0393}$ & 0.3853$^{\pm .0486}$ \\
      & \textit{w/o} RES & 0.5167$^{\pm .0251}$ & 0.4150$^{\pm .0283}$ \\
    \bottomrule
  \end{tabular}}
  \label{tab:Ablation-single}
\end{table}

\begin{table}[ht]
  \centering
  \caption{Ablation study of \textit{SCORE} in multi-agent settings across three scenes.}
  \begin{tabular*}{0.475\textwidth}{@{\extracolsep{\fill}}llc}
    \toprule
    Scenario & Method & DTW $\downarrow$ \\
    \midrule
    \multirow{4}{*}{Zara01}
      & \textit{SCORE} & \textbf{0.3616}$^{\pm .1015}$ \\
      & \textit{w/o all Motivational Drivers} & 0.4721$^{\pm .0929}$ \\
      & \textit{w/o LLM} & 0.4711$^{\pm .0939}$ \\
      & \textit{w/o RES} & 0.5193$^{\pm .0840}$ \\
    \midrule
    \multirow{4}{*}{Zara02}
      & \textit{SCORE} & \textbf{0.2657}$^{\pm .0832}$ \\
      & \textit{w/o all Motivational Drivers} & 0.3021$^{\pm .0730}$ \\
      & \textit{w/o LLM} & 0.3125$^{\pm .0719}$ \\
      & \textit{w/o RES} & 0.4116$^{\pm .0651}$ \\
    \midrule
    \multirow{4}{*}{student003}
      & \textit{SCORE} & \textbf{0.4471}$^{\pm .0956}$ \\
      & \textit{w/o all Motivational Drivers} & 0.5071$^{\pm .0881}$ \\
      & \textit{w/o LLM} & 0.6192$^{\pm .0718}$ \\
      & \textit{w/o RES} & 0.6102$^{\pm .0652}$ \\
    \bottomrule
  \end{tabular*}
  \label{tab:DTW_Ablation-multi}
\end{table}


\begin{figure}[t]
  \centering
  \includegraphics[width=\linewidth]{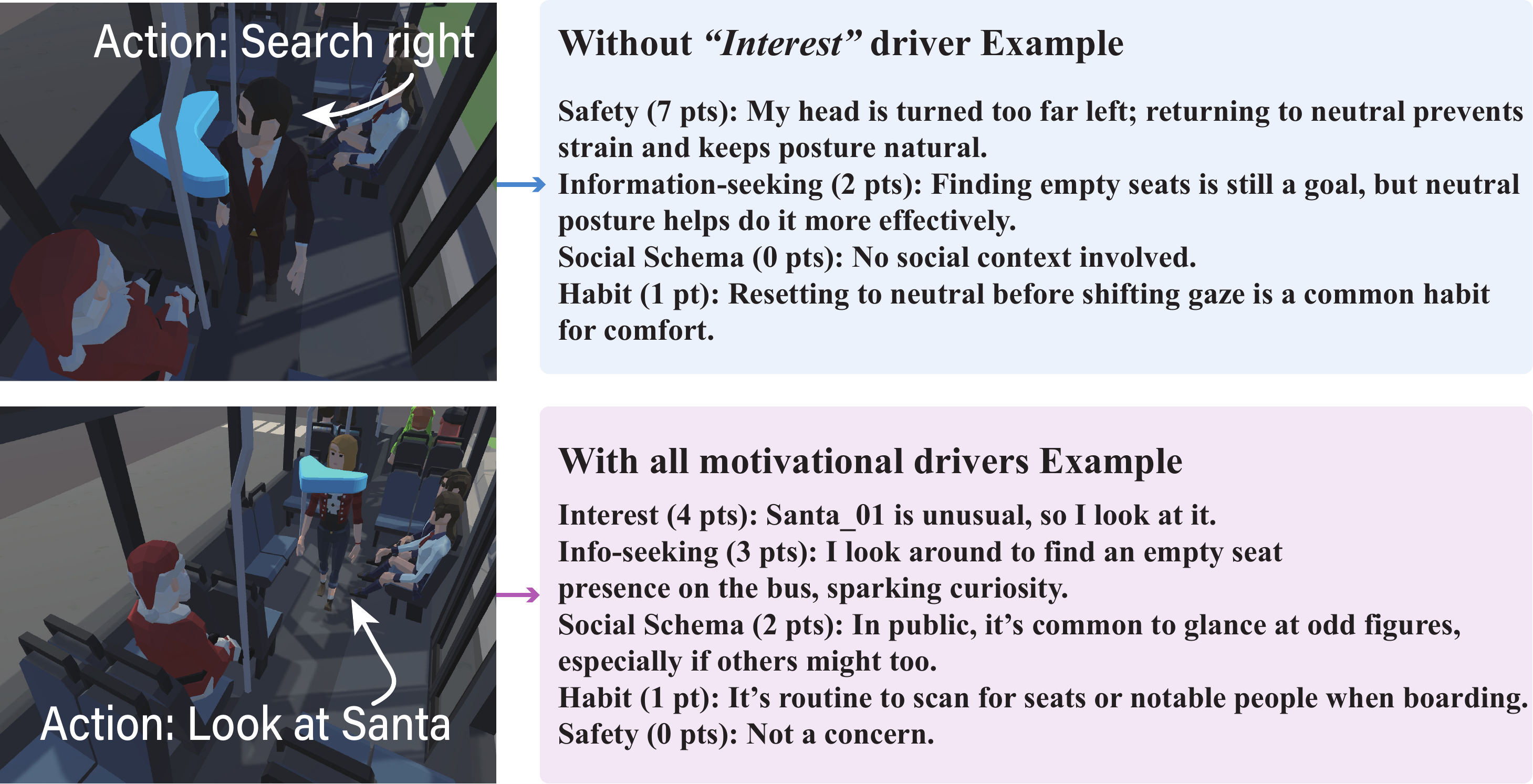}
  \caption{Visualization of action and reasoning result when \textit{Interest} is removed from Motivational Drivers.}
  \label{fig:NoInterestAblation_Visualization}
\end{figure}

\subsection{Responsiveness to Unseen Distractors}
To evaluate the responsiveness enabled by SCORE's semi-online RES, we compare model-generated head trajectories under three conditions against \textit{Human} recorded in the APC, using DTW as the similarity metric. The three experimental conditions are defined as follows:

\begin{sloppypar}
\begin{itemize}
\item \textit{MDC-MDC}: Both DPS and RES are run in the MDC. As the model is never exposed to the added distractors present in the APC, this condition serves as a lower bound on performance.
\item \textit{MDC-APC}: DPS is executed in the MDC, while RES runs in the APC. Any improvement over \textit{MDC-MDC} reflects RES’s ability to detect and adapt to novel stimuli not available at planning time. 
\item \textit{APC-APC}: Both DPS and RES run in the APC, giving the model full access to distractors. This represents an upper bound on performance.
\end{itemize}
\end{sloppypar}

Results are reported in~\autoref{tab:DTW_Responsiveness Benchmark}. Notably, \textit{MDC-APC} yields an improvement over \textit{MDC-MDC}, indicating that semi-online validation not only mitigates hallucinations but also enhances responsiveness to scene changes introduced at runtime. 
Qualitative footage in the supplementary video further illustrates timely reactions to suddenly appearing events, such as an unexpected Santa Claus crossing the field of view.

\subsection{Ablation Study}
We performed an ablation study to isolate the contribution of \textit{SCORE}’s core components.
First, we removed the five symbolic cognitive drivers from the prompts to test whether explicit motivational drivers improve plausibility. All five drivers were removed from the prompt inputs (\textit{w/o all Motivational Drivers}). Furthermore, each driver was selectively excluded while keeping the others intact (\textit{w/o Interest driver}, \textit{w/o Information Seeking driver}, \textit{w/o Safety driver}, \textit{w/o Social Schema driver}, or \textit{w/o Habit driver}).
Second, we disabled the LLM and allowed the VLM to choose head poses directly (\textit{w/o LLM}) to evaluate the benefit of a dedicated planning stage.
Third, we replaced the deliberative–reactive pipeline with a purely deliberative pass (\textit{w/o RES}) to determine whether the lightweight validation step truly mitigates hallucinations. 

We conducted ablation studies in both single- and multi-agent scenarios, with different emphases in terms of motivational drivers. In the single-agent setting, we performed both full ablations. In the multi-agent setting, we limited ablations to the complete removal of all drivers, omitting individual factor tests. This decision was based on two factors: the single-agent results already demonstrate that partial ablations fall between the full model and no-driver baseline, and repeating this in the multi-agent case would add redundancy without yielding new insights.

Results are summarized in~\autoref{tab:Ablation-single} (single-agent) and ~\autoref{tab:DTW_Ablation-multi} (multi-agent). Removing cognitive drivers increases DTW error in all scenes, confirming that explicit driver modelling is essential for human-like reasoning. 
Eliminating the LLM and delegating planning entirely to the VLM degrades performance, indicating that a dedicated reasoning stage is essential for producing coherent head trajectories.
Finally, omitting the RES amplifies hallucination-induced errors and yields poor scores, demonstrating that FastVLM validation is critical for robust, believable motion. 
The same trends hold when many agents operate simultaneously, indicating that each component scales effectively to group scenarios.

\begin{figure*}[t]
  \centering
  \includegraphics[width=1\linewidth]{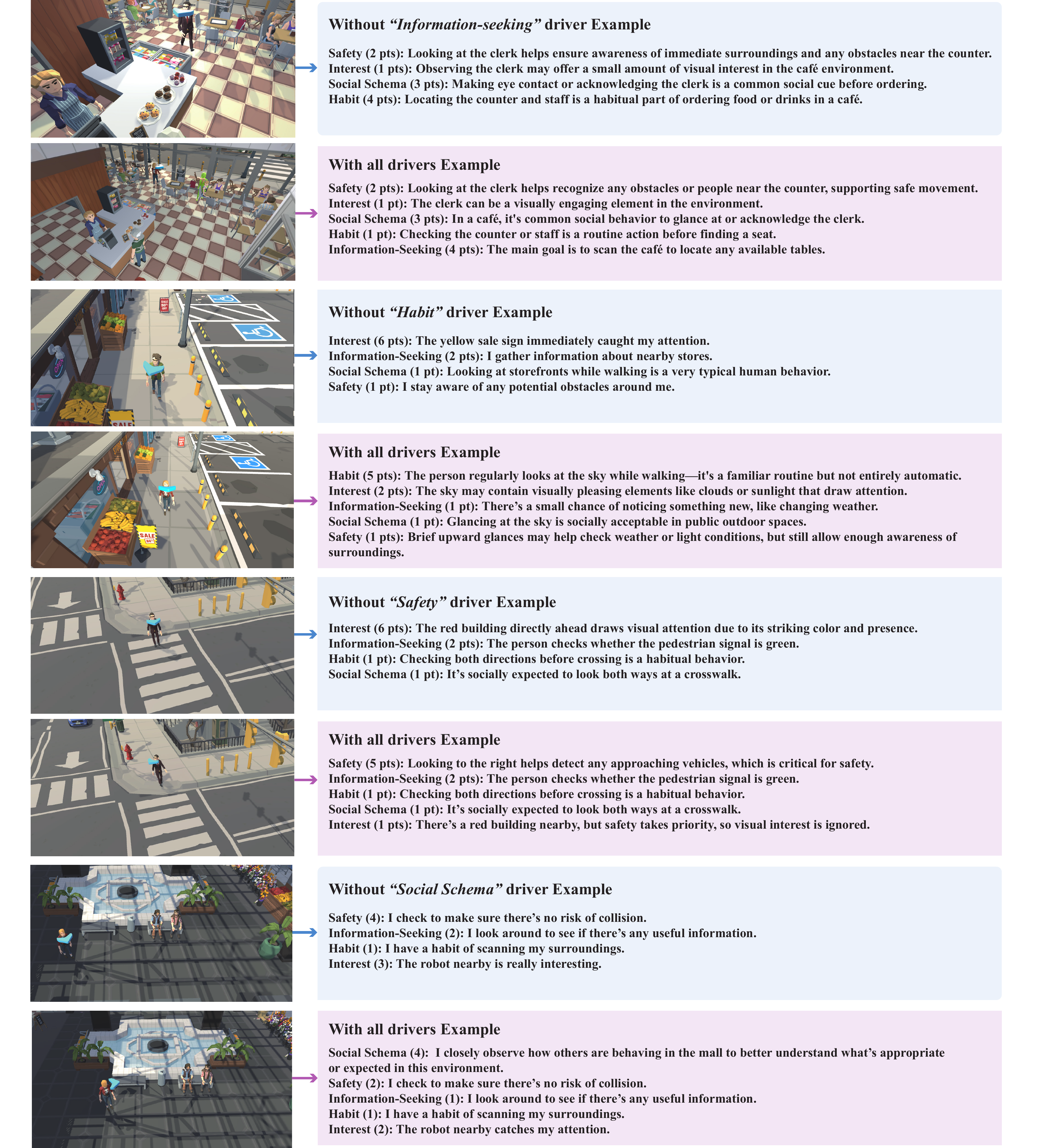}
  \caption{Visualization of reasoning and action when \textit{Information-seeking},\textit{Habit},\textit{Safety} and \textit{Social schema} are excluded from the motivational drivers.}
  \label{fig:RemovedMotivationalDrivers}
\end{figure*}

We additionally visualized how the design of the five factors in our Motivational Drivers impacts the model's reasoning. As shown in ~\autoref{fig:NoInterestAblation_Visualization}, when the \textit{Interest} factor is removed, the agent does not look at Santa even though it is interesting and atypical in the bus. 
The analysis of other factors can be found in ~\autoref{fig:RemovedMotivationalDrivers}.

\section{Conclusion and Future works}
This paper presents \textit{SCORE}, a language-guided framework that couples vision-language perception with language-based reasoning to synthesize human-like head motion. Built on a controlled VR user study, \textit{SCORE} embeds five empirically derived cognitive drivers—\textit{Interest}, \textit{Information-seeking}, \textit{Safety}, \textit{Social schema}, and \textit{Habit}—into a two-stage pipeline.
An offline deliberative stage generates a time-stamped head-orientation plan, and an online reactive stage verifies each pose with a lightweight FastVLM, suppressing hallucinations and adapting to late scene changes.
Across single- and multi-agent settings, this design more closely replicates human trajectories than the saliency baseline Track, underscoring the value of language-guided cognition for context-aware gaze control.

We found that \textit{SCORE}’s rotation decision occasionally diverges from that of humans, reflecting differences not in error but in individual tendencies or persona-like styles. Even under identical conditions, participants displayed substantial variability in where and why they turned their heads, meaning that only probabilistic distributions of plausible behaviors can be inferred rather than a single ground truth trajectory. Similarly, SCORE sometimes produced conclusions that differed from human responses but were still reasonable—often reflecting a more safety-conscious or ``diligent” strategy. These results suggest that aligning with human diversity may require modeling variability itself rather than assuming uniformity.

Looking ahead, \textit{SCORE} holds strong potential for broader applications through further refinement and integration with complementary technologies. First, the current system relies solely on visual inputs, overlooking richer perceptual modalities. Incorporating spatial audio and haptic feedback would enable agents to respond to non-visual cues such as approaching footsteps or vibration alerts. Second, the framework presently operates independently of locomotion control. Coupling \textit{SCORE} with a model-predictive navigation controller could unify locomotion and gaze, enabling coordinated whole-body behaviors in constrained or collaborative tasks. Third, reliance on large-scale vision–language models introduces inference latency. Distilling the LLM into a task-specific lightweight model and caching frequently used predicate patterns may reduce latency while retaining reasoning depth.
Pursuing these extensions could extend \textit{SCORE} to large-scale social VR, robot telepresence, and other applications that demand believable, responsive head motion.

\begin{acks}
This work was supported by the \grantsponsor{GSNRF}{National Research Foundation of Korea}{https://www.nrf.re.kr/eng/main} under Grant No.~\grantnum{GSNRF}{RS-2025-00518643} (34\%). It was also partly supported by ICT Creative Consilience Program through the Institute of Information \& Communications Technology Planning \& Evaluation (IITP) grant funded by the Korea government (MSIT) under Grant No. IITP-2025-RS-2020-II201819 (33\%). It was also supported by Institute of Information \& communications Technology Planning \& Evaluation  (IITP) grant funded by the Korea government (MSIT) High-Performance Research AI Computing Infrastructure Support at the 2 PFLOPS Scale under Grant No.RS-2025-02653113 (33\%).
\end{acks}

\clearpage
\bibliographystyle{ACM-Reference-Format}
\bibliography{references}





\end{document}